\documentclass[11pt]{article}

\usepackage{amsmath}
\usepackage{epsfig}
\usepackage{graphics}

\newcommand{\N}{N\raise.7ex\hbox{\underline{$\circ $}}$\;$}

\begin{document}

\title{
 V.M. Red'kov\footnote{redkov@dragon.bas-net.by}, N.G. Tokarevskaya , and V.V. Kisel\\
Graviton in a Curved Space-Time Background\\ and Gauge Symmetry }

\maketitle

\begin{quotation}

Pauli-Fierz approach to description of a massless spin-2 particle
is investigated in the framework of 30-component first order
relativistic wave equation theory on a curved space-time
background. It is shown that additional gauge symmetry of massless
equations established by Pauli-Fierz can be  extended only  to
curved space-time regions where  Ricci tensor vanishes. In all such
space-time models the  generally covariant S=2 massless wave
equation exhibits gauge symmetry  property, otherwise it is not
so.

\end{quotation}

\section{Introduction
}


The theory  of the massive spin-2 field has received much attention over the years
since  the initial construction of  Lagrangian formulation by Fierz and Pauli
[1-2].
The original Fierz-Pauli theory  for spin  was second order in derivatives $\partial_{\alpha}$
(and involved scalar and tensor auxiliary fields).
It is  highly  satisfactory as long  as we restrict ourselves to  a free particle case.
However this approach turned out not to be so good at considering spin-2 theory in presence of an external
electromagnetic field.  Federbush [3] showed that  to avoid  a loss of constrains problem the minimal
coupling had to be  supplemented by a direct non-minimal to the electromagnetic field strength.
There followed a number of works on modification or  generalizations  of the Fierz-Pauli theory
(Rivers [4], Nath [5], Bhargava and Watanabe [6], Tait [7], Reilly [8]).
At the same time  interest in  general high-spin  fields  was generated by  the discovery of the now
well-known inconsistency problems of Johnson and Sudarshan [9] and  Velo and Zwanzinger [10].
In the course of  investigating  their acausality problems for other then 3/2, Velo-Zwanzinger
rediscovered  the spin-2 loss of constrains problem, but were not at first aware of  the non-minimal
 term solution of it.
 Velo [11] later made  a  thorough analysis of the external field  problem
for  the 'correct'  non-minimally coupled  spin-2 theory, showing  that it   is acausal too.

All the work mentioned  above dealt  with  a second-order formalism for the spin-2 theory. Much of the
confusion  which arose over this theory  could be traced to  the so-called "derivative ordering ambiguity
(Naglal [12]). This problem  can be avoided by working from the start with a first-order  formalism
(for example see Gel'fand et al [13]) and for which  the minimal coupling procedure is unambiguous..

The work by Fedorov [14] was likely to be the first one where consistent investigation of
the spin-2 theory in the framework of first-order theory was carried out in detail.
The 30-component wave equation  [14] referred to the so-called canonical basis,
transition from which to the more familiar tensor formulation is possible but laborious task and
it was not done in
{[14].
Subsequently the same 30-component theory  was rediscovered and  fundamentally elaborated
in tensor-based approach by a number of authors (Regge [15], Schwinger [16],  Chang [17], Hagen [18],
 Mathews et al [19], Cox [20]). Also a matrix  formalism for the spin-2 theory was developed
 (Fedorod, Bogush, Krylov, Kisel  [21-25]).

Concurrently else one theory for spin-2 particle was advanced that requires 50 field  components
(Adler [26],  Deser et al [27], Fedorov and Krylov  [28, 23], Cox [20.]). It appears to be more
complicated, however some evident  correlation between  the corresponding massless theory  and
the non-linear gravitational  equation is revealed (Fedorov [28]).

Possible connections between two variants of spin-2 theories  have
been investigated. Seemingly, the most  clarity was achieved by
Bogush and Kisel [25], who showed that  50-component equation in
presence of an external electromagnetic field can be reduced to 30-component  equation
with additional interaction  that must be interpreted as anomalous magnetic momentum term.

In the present work the 30-component first order theory is investigated in the case of vanishing mass
of  the particle and external curved space-time background.

\section{Particle in the flat space-time}

A  system of first order wave equations describing a massless spin-2 particle in
a flat space-time has the form
\begin{eqnarray}
\partial^{a} \Phi _{a} = 0 \; ,
\label{1.1a}
\\
{1 \over 2}  \partial_{a} \Phi  -  {1
\over 3}  \partial^{b} \Phi_{ab}  =  \Phi_{a}  ,
\label{1.1b}
\\
{1 \over 2} ( \partial^{k} \Phi_{kab}  +
\partial^{k} \Phi_{kba} - {1 \over 2}  g_{ab}
\partial^{k} \Phi _{kn}^{\;\;\;\;n}  )
+ \partial_{a} \Phi_{b} +  \partial_{b} \Phi_{a}-{1
\over 2} g_{ab}   \partial^{k} \Phi_{k}  = 0   ,
\\
\partial_{a} \Phi_{bc} - \partial_{b} \Phi_{ac}
 + {1 \over 3}
(g_{bc}  \partial^{k} \Phi_{ak} -   g_{ac}  \partial^{k}
\Phi_{bk} )  =  \Phi _{abc} \; .
\label{1.1d}
\end{eqnarray}

\noindent
A 30-component wave function consists of
a scalar $\Phi$,   vector  $\Phi_{a}$, symmetric 2-rank tensor  $\Phi_{ab}$,
and  3-rank tensor $\Phi _{abc}$ antisymmetric in two first indices.
From  (\ref{1.1d}) it follows four conditions that are satisfied by the 3-index field:
\begin{eqnarray}
\Phi_{abc}  +  \Phi_{bca}  +  \Phi_{cab} = 0
\mbox{ or} \;\;\; \epsilon^{kabc}   \Phi_{abc}  = 0  .
\label{1.2a}
\end{eqnarray}

\noindent Simplifying Eq.  (\ref{1.1d}) in indices  $b$ and  $c$, one  produces
\begin{eqnarray}
\partial_{a}  \Phi^{\;\;b}_{b} = \Phi_{ac}^{\;\;\;\;c} \; .
\label{1.2b}
\end{eqnarray}

\noindent
Thus, a total number of independent components entering the theory  equals 31
(instead of 30 in massive case):
\begin{eqnarray}
\Phi
(x)  \Longrightarrow  1  ,
\qquad
\Phi_{a}
\Longrightarrow  4  ,
\qquad
\Phi_{ab} \Longrightarrow
10   ,
\nonumber
\\
\Phi_{abc}  \Longrightarrow
6 \times 4 - 4 -4  = 16  .
\nonumber
\end{eqnarray}

After excluding  fields $\Phi_{a}$ and $\Phi_{kab}$ from (\ref{1.1a}-\ref{1.1d})
one gets to a pair of second order equations on fields  $\Phi (x) $  and  $\Phi_{ab}(x)$\footnote{
The notation $\nabla^{2} = \partial^{a} \partial_{a}$ is used.}:
\begin{eqnarray}
{1 \over 2}  \nabla^{2} \Phi
- {1 \over 3}  \partial^{k} \partial^{l} \Phi_{kl} = 0,
\label{1.3a}
\\
(\partial_{a} \partial_{b} -   {1
\over 4} g_{ab} \nabla^{2}  )  \Phi
- {1 \over 4}
g_{ab}   \nabla^{2}  \Phi^{c}_{c}
+   \nabla^{2}  \Phi_{ab}
-   \partial_{a} \partial^{l}  \Phi_{bl}
 -\partial_{b} \partial^{l}  \Phi_{al}  +
 {1 \over 2} g_{ab}  \partial^{k} \partial^{l}  \Phi_{kl} = 0  .
\label{1.3b}
\end{eqnarray}

\noindent Allowing for  (\ref{1.3a}), Eq. (\ref{1.3b}) can be rewritten as
\begin{eqnarray}
( \partial_{a} \partial_{b}  +   {1
\over 2} g_{ab} \nabla^{2}  )  \Phi  -  {1 \over 4}
g_{ab}   \nabla^{2}  \Phi^{c}_{c}
 +   \nabla^{2}  \Phi_{ab}
 -
\partial_{a} \partial^{l}  \Phi_{bl}  -
 \partial_{b} \partial^{l}  \Phi_{al}   =  0  .
\label{1.3c}
\end{eqnarray}

The fact of prime significance in the  theory under consideration is that
these equations permit specific gauge principle \footnote{The fact was firstly established
by Pauli and Fierz  [1-2] .}. That means the following:
 the above second order system  (\ref{1.3c}) is satisfied by a
a substitution (class of trivial or gradient-like solution)
\begin{eqnarray}
\Phi^{(0)} = \partial^{l} \Lambda_{l}  ,
\Phi^{(0)} _{ab}  =
\partial_{a} \Lambda_{b}  +
\partial_{b} \Lambda_{a}  - {1 \over 2}  g_{ab}  \partial^{l} \Lambda_{l}  ,
\label{1.4}
\end{eqnarray}

\noindent at any 4-vector function $\Lambda_{a}(x)$. Indeed,
\begin{eqnarray}
 -{1 \over 3} \;   \partial^{a} \partial^{b}
\Phi^{(0)}_{ab} = - {1 \over 2}  \nabla^{2}  \partial^{l}
\Lambda_{l} = - {1 \over 2}  \nabla^{2}  \Phi^{(0)}   ,
\label{1.5}
\end{eqnarray}

\noindent and therefore the set (\ref{1.4})  turns Eq. (\ref{1.3a}) into identity. Further,
taking into account
\begin{eqnarray}
{1 \over 2}  ( \partial^{k} \Phi_{kab}  +
\partial^{k} \Phi_{kba}  - {1 \over 2}  g_{ab}
\partial^{k} \Phi _{kn}^{\;\;\;\;n}  )
  =+{1 \over 3} \partial^{l}\partial_{l} (\partial_{b} \Lambda_{a} + \partial_{a} \Lambda_{b})-
 {2 \over 3}  \partial_{a} \partial_{b}  \partial^{l}\Lambda_{l}  ,
\nonumber
\\
 \partial_{a} \Phi_{b}  +   \partial_{b} \Phi_{a} -  {1
\over 2} g_{ab}   \partial^{k} \Phi_{k}
= -{1 \over 3} \partial^{l}\partial_{l} (\partial_{b} \Lambda_{a} + \partial_{a} \Lambda_{b})+
 {2 \over 3} \; \partial_{a} \partial_{b} \; \partial^{l}\Lambda_{l}  ,
\nonumber
\end{eqnarray}

\noindent one can verify that the set  (\ref{1.4}) satisfies Eq.  (\ref{1.3b}) as well.

So, a massless spin-2 field in Minkowski space-time can be  described by the first order
system, or by the second order system (Pauli-Fierz [1-2]).
At this their solutions are  not determined uniquely; in general, to any chosen one
we  may add an arbitrary $\Lambda_{a}$ -dependent term.

\section{ Particle in curved space-time }

With the use of principle of minimal coupling to a curved space-time background
(external gravitational field), expected  generally covariant equations for a spin-2
particle are to be taken in the form
\begin{eqnarray}
\nabla^{\alpha}  \Phi _{\alpha} = 0  , \qquad \qquad \qquad
\label{2.1a}
\\[0.2cm]
 {1 \over 2}  \nabla_{\alpha}  \Phi  -  {1 \over 3}  \nabla^{\beta}
\Phi_{\alpha \beta  } =   \Phi_{\alpha}  , \qquad
\label{2.1b}
\\[0.2cm]
{1 \over 2}  \Big( \nabla^{\rho }  \Phi_{\rho  \alpha \beta }  +
\nabla^{\rho }  \Phi_{\rho  \beta  \alpha }  -
  {1 \over 2}
g_{\alpha  \beta}(x)   \nabla^{\rho}  \Phi_{\rho
\sigma}^{\;\;\;\;\sigma }\; \Big)
+ \Big( \; \nabla_{\alpha} \Phi_{\beta} + \nabla_{\beta} \Phi_{\alpha} -
{1 \over 2} g_{\alpha \beta }(x)  \nabla^{\rho }  \Phi_{\rho} \Big)  =
0  ,
\label{2.1c}
\\[0.2cm]
\nabla_{\alpha} \Phi_{\beta\sigma }  - \nabla_{\beta} \Phi_{\alpha  \sigma}
+ {1 \over 3}  (g_{\beta  \sigma  }(x)  \nabla^{\rho}  \Phi_{\alpha \rho  }
- g_{\alpha \sigma}(x) \nabla^{\rho } \Phi_{\beta \rho} )  =  \Phi _{\alpha \beta \sigma}  .
\qquad \qquad
\label{2.1d}
\end{eqnarray}

\noindent Here  $\nabla_{\alpha}$ designates a  generally covariant derivative.
As in the flat space-time,  the system exhibits  the property
\begin{eqnarray}
\nabla_{\alpha} \; \Phi^{\;\;\beta}_{\beta} = \Phi_{\alpha \beta }^{\;\;\;\;\beta } \; .
\label{2.2}
\end{eqnarray}

Now we are to investigate the question of  possible gauge symmetry of the system.
To this end we will try to satisfy these equations  by a substitution
\begin{eqnarray}
\Phi^{(0)} = \nabla^{\beta}  \Lambda_{\beta}   ,  \qquad \qquad \qquad
\nonumber
\\
\Phi^{(0)} _{\alpha \beta}  =
\nabla _{\alpha}  \Lambda_{\beta}   +
\nabla_{\beta}  \Lambda_{\alpha}    -
{1 \over 2} g_{\alpha \beta }(x)  \nabla^{\sigma} \Lambda_{\sigma}  ,
\label{2.3}
\end{eqnarray}

\noindent where  $\Lambda (x)$ is an arbitrary 4-vector function.
With the use of Eq. (\ref{2.1b}), a vector field corresponding to the set (\ref{2.3})
takes the form
\begin{eqnarray}
\Phi^{(0)}_{\alpha} =
{2 \over 3} \nabla_{\alpha} \nabla^{\beta}  \Lambda_{\beta}  -
{1 \over 3} \nabla^{\beta}  \nabla_{\alpha} \Lambda_{\beta}
 -
{1 \over 3} (\nabla^{\beta} \nabla_{\beta}) \Lambda_{\alpha}  .
\label{2.4}
\end{eqnarray}

\noindent After substitution it into Eq. (\ref{2.1a}) one produces
\begin{eqnarray}
0 =
{2 \over 3}( \nabla^{\alpha}   \nabla_{\alpha} ) \nabla^{\beta}  \Lambda_{\beta}  -
{1 \over 3}  \nabla^{\alpha}  \nabla^{\beta}  \nabla_{\alpha}  \Lambda_{\beta}
- {1 \over 3}  \nabla^{\beta}  (\nabla^{\alpha}  \nabla_{\alpha})  \Lambda_{\beta}  . \qquad \qquad
\label{2.5a}
\end{eqnarray}

\noindent Employing  conventionally the Riemann and Ricci tensors
\begin{eqnarray}
(\nabla_{\beta}  \nabla_{\alpha} - \nabla_{\alpha}  \nabla_{\beta} ) \Lambda_{\rho} =
R_{\beta \alpha \rho \sigma}  \Lambda^{\sigma}  , \;
R_{\beta \alpha ... \sigma}^{\;\;\;\;\;\beta}   = R_{\alpha \sigma} ,
\nonumber
\end{eqnarray}

\noindent the second term in (\ref{2.5a}) can be rewritten as
\begin{eqnarray}
-{1 \over 3}  \nabla^{\alpha} \nabla_{\beta}  \nabla_{\alpha}  \Lambda^{\beta}  =
-{1 \over 3}  \nabla^{\alpha} (  \nabla_{\alpha }  \nabla_{\beta} \Lambda^{\beta}
+  R_{\alpha \beta}  \Lambda^{\beta}  )  ,
\nonumber
\end{eqnarray}

\noindent with the use of which Eq. (\ref{2.5a}) will take the form
\begin{eqnarray}
0 = {1 \over 3}   [\nabla^{\alpha} \nabla_{\alpha} ,  \nabla^{\beta} ]_{-}
\Lambda_{\beta}    -
{1 \over 3} \nabla^{\alpha} ( R_{\alpha \beta} \Lambda^{\beta}  ) .
\label{2.5b}
\end{eqnarray}

\noindent
The latter, with  the commutator
\begin{eqnarray}
 [ \nabla^{\alpha} \nabla_{\alpha},     \nabla^{\beta} ]_{-}  \Lambda_{\beta}
 =-  \nabla^{\alpha} (R_{\alpha \sigma}  \Lambda^{\sigma})  ,
\label{2.6}
\end{eqnarray}

\noindent will read as
\begin{eqnarray}
0 = -{2 \over 3}  \nabla^{\alpha} (R_{\alpha \beta} \Lambda^{\beta} )  .
\label{2.7}
\end{eqnarray}

 This equation means:  if $R_{\alpha\beta} \neq 0$,  the
present spin-2 particle equations do not have any trivial
$\lambda_{a}$-based solution. In other terms, a gauge principle in
accordance with Einstein gravitational equations the equality
$R_{\alpha\beta} \neq 0$ speaks that  at those $x^{\alpha}$-points
any material  fields vanish. However, in $(R_{\alpha\beta} = 0)$
-region the  wave equation under consideration includes such
$\lambda_{\alpha}$-based solutions and
correspondingly a gauge principle. Now, analogously, we should
consider Eq. (\ref{2.1c}): what will we have had on substituting
$\Lambda_{\alpha}$-set into it. We must exclude all auxiliary
fields from Eq.~(\ref{2.1c}):
\begin{eqnarray}
{1 \over 2}  ( \nabla^{\rho }  \Phi^{(0)}_{\rho  \alpha \beta }  +
\nabla^{\rho }  \Phi^{(0)}_{\rho  \beta  \alpha } -   {1 \over 2}
g_{\alpha  \beta}(x)   \nabla^{\rho}  \Phi^{(0)\;\;\;\;\sigma}_{\rho
\sigma}\; )
\nonumber \\
+ \nabla_{\alpha} \Phi^{(0)}_{\beta}  + \nabla_{\beta} \Phi^{(0)}_{\alpha} -
{1 \over 2} g_{\alpha \beta }(x)  \nabla^{\rho }\Phi^{(0)}_{\rho}   =
0  ,
\label{2.8}
\end{eqnarray}

\noindent
Let us step by step calculate all terms entering Eq.~(\ref{2.8}). For first (1)  term we  have that
\begin{eqnarray}
(1) \stackrel{def}{=\hspace{-0.1cm}=}
{1 \over 2}   \nabla^{\rho }  \Phi^{(0)}_{\rho  \alpha \beta }
 =
{1 \over 2} ( \nabla^{\rho} \nabla_{\rho} ) (\nabla_{\alpha} \Lambda_{\beta})
\nonumber
\\
+ {1 \over 2} ( \nabla^{\rho} \nabla_{\rho} ) (\nabla_{\beta} \Lambda_{\alpha}) -
{1\over 4} g_{\alpha \beta} (\nabla^{\rho}\nabla_{\rho})(\nabla^{\gamma} \Lambda_{\gamma})
\nonumber \\
- {1 \over 2}  (\nabla^{\rho} \nabla_{\rho } )\nabla_{\alpha}   \Lambda_{\beta}
-
{1 \over 2}  \nabla^{\rho} [\nabla_{\alpha} , \nabla_{\rho }]_{-} \Lambda_{\beta}
\nonumber
\\
-
{1 \over 2}   \nabla_{\alpha}  \nabla_{\beta}  (\nabla^{\rho}\Lambda_{\rho}) -
{1 \over 2}  [\nabla^{\rho}, \nabla_{\alpha}  \nabla_{\beta}]_{-} \Lambda_{\rho}
\nonumber \\
+ {1\over 4}  (\nabla_{\beta}\nabla_{\alpha})(\nabla^{\gamma}
\Lambda_{\gamma})
 + {1 \over 6} g_{\alpha \beta}
(\nabla^{\rho} \nabla_{\rho} )( \nabla^{\sigma} \Lambda_{\sigma})
\nonumber \\[0.2cm]
+
  {1 \over 6} g_{\alpha \beta} \nabla^{\rho} [\nabla^{\sigma}, \nabla_{\rho} ]_{-} \Lambda_{\sigma}
+
{1 \over 6} g_{\alpha \beta}  (\nabla^{\sigma}\nabla_{\sigma})  (\nabla^{\rho}\Lambda_{\rho})
\nonumber
\\
 +
{1 \over 6} g_{\alpha \beta} [\nabla^{\rho}, \nabla^{\sigma}\nabla_{\sigma} ]_{-} \Lambda_{\rho}
-
{1 \over 12}  g_{\alpha \beta} (\nabla^{\rho}\nabla_{\rho}) (\nabla^{\gamma} \Lambda_{\gamma})
\nonumber \\
-
{1 \over 6} \nabla_{\beta}  \nabla_{\alpha} (\nabla^{\sigma} \Lambda_{\sigma}) -
{1 \over 6} \nabla_{\beta} [\nabla^{\sigma} , \nabla_{\alpha}]_{-} \Lambda_{\sigma}
\nonumber
\\
-
{1 \over 6}  (\nabla^{\sigma}\nabla_{\sigma}) \nabla_{\beta}  \Lambda_{\alpha} -
{1 \over 6} [\nabla_{\beta} , \nabla^{\sigma}\nabla_{\sigma} ]_{-} \Lambda_{\alpha}
+ {1 \over 12} \nabla_{\beta} \nabla_{\alpha} (\nabla^{\gamma} \Lambda_{\gamma}) .
\nonumber
\end{eqnarray}

\noindent
Second term in  Eq.~(\ref{2.8}) can be produced on  straightforward symmetry considerations from
Eq.~(\ref{2.8}).
Third term in  Eq.~(\ref{2.8})  turns out to vanish
\begin{eqnarray}
(3) \stackrel{def}{=\hspace{-0.1cm}=}-{1 \over 4} g_{\alpha \beta} \nabla^{\rho} \Phi^{(0)\;\;\gamma}_{\rho \gamma}
= -{1 \over 2} g_{\alpha \beta} \nabla^{\rho} \nabla_{\rho} \Phi^{(0)\beta}_{\beta}
 \nonumber \\
=
-{1 \over 4} g_{\alpha \beta} \nabla^{\rho} \nabla_{\rho}\; (\nabla_{\beta} \Lambda^{\beta} +
 \nabla_{\beta} \Lambda^{\beta}  -
 {1 \over 2} \delta^{\beta}_{\beta} \; \nabla^{\gamma}\Lambda_{\gamma}) = 0 .
\nonumber
\end{eqnarray}

\noindent
For fourth and fifth terms we  will have

\begin{eqnarray}
(4) \stackrel{def}{=\hspace{-0.1cm}=}
\nabla_{\alpha} \; \Phi^{(0)}_{\beta}
=
{1 \over 2} \nabla_{\alpha} \nabla_{\beta} \; \nabla^{\gamma} \Lambda_{\gamma} \;-\;
  {1 \over 3} \;\nabla_{\alpha}  \nabla_{\beta} (\nabla^{\rho} \;   \Lambda_{\rho})
\nonumber\\
-  {1 \over 3} \;\nabla_{\alpha}  [\nabla^{\rho} ,  \nabla_{\beta} ]_{-} \Lambda_{\rho} -
  {1 \over 3} \; ( \nabla^{\rho} \; \nabla_{\rho} ) \;  \nabla_{\alpha}   \Lambda_{\beta}
\nonumber
\\[0.2cm]
-  {1 \over 3} \; [\nabla_{\alpha} ,  \nabla^{\rho} \; \nabla_{\rho}]_{-} \Lambda_{\beta} +
 {1 \over 6} \;
\nabla_{\alpha}  \nabla_{\beta} \;  \nabla^{\gamma} \Lambda_{\gamma} \; ,
\nonumber
\label{2.11a}
\\[0.2cm]
(5) \stackrel{def}{=\hspace{-0.1cm}=}
\nabla_{\beta} \; \Phi^{(0)}_{\alpha}  =
 {1 \over 2} \nabla_{\beta} \nabla_{\alpha} \; \nabla^{\gamma} \Lambda_{\gamma}
\nonumber\\
-  {1 \over 3} \;\nabla_{\beta}  \nabla_{\alpha} (\nabla^{\rho} \;   \Lambda_{\rho})
-  {1 \over 3} \;\nabla_{\beta}  [\nabla^{\rho} ,  \nabla_{\alpha} ]_{-} \Lambda_{\rho}
-  {1 \over 3} \; ( \nabla^{\rho} \; \nabla_{\rho} ) \;  \nabla_{\beta}   \Lambda_{\alpha}
\nonumber
\\[0.2cm]
-  {1 \over 3} \; [\nabla_{\beta} ,  \nabla^{\rho} \; \nabla_{\rho}]_{-} \Lambda_{\alpha}
+  {1 \over 6} \;
\nabla_{\beta}  \nabla_{\alpha} \;  \nabla^{\gamma} \Lambda_{\gamma} ;
\nonumber
\end{eqnarray}

\noindent and term (6) is
\begin{eqnarray}
(6) \stackrel{def}{=\hspace{-0.1cm}=} -{1 \over 2} g_{\alpha \beta}  \nabla^{\rho} \Phi^{(0)}_{\rho}
  -{1 \over 2} g_{\alpha \beta}  \nabla^{\rho} \Phi^{(0)}_{\rho}
\nonumber\\
=
-{1 \over 4} g_{\alpha \beta} (\nabla^{\rho} \nabla_{\rho})
 (\nabla^{\gamma}\Lambda_{\gamma})+
 {1 \over 6}  g_{\alpha \beta} (\nabla^{\rho}
\nabla_{\rho}) (\nabla^{\sigma}  \Lambda_{\sigma})
\nonumber
\\[0.2cm]
+ {1 \over 6}
g_{\alpha \beta} \nabla^{\rho} [\nabla^{\sigma} , \;\nabla_{\rho}]_{-}
 \Lambda_{\sigma} +
{1 \over 6} g_{\alpha \beta} (\nabla^{\sigma} \;
\nabla_{\sigma} ) (\nabla^{\rho}  \Lambda_{\rho} ) \nonumber\\
+ {1 \over 6} g_{\alpha \beta} [\nabla^{\rho},
\nabla^{\sigma} \nabla_{\sigma}]_{-} \Lambda_{\rho} -
{1 \over 12} g_{\alpha \beta} (\nabla^{\rho} \nabla_{\rho}) (\nabla^{\gamma} \Lambda_{\gamma}) .
\nonumber
\end{eqnarray}

\noindent Summing up all six expressions and taking into account similar terms
(factors at all terms without commutators  turn out to be equal zero as should be
expected):
\begin{eqnarray}
0 = (\nabla^{\rho}\nabla_{\rho})\; (\nabla_{\alpha}\Lambda_{\beta})
\Big[ \Big({1\over 2} - {1\over 2}\Big)  +
\Big({1 \over 2} - {1 \over 6}\Big)  -   {1\over 3}  \Big]
\nonumber \\
+  (\nabla^{\rho}\nabla_{\rho})  (\nabla_{\beta}\Lambda_{\alpha})
\Big[  \Big({1\over 2} - {1\over 6}\Big)   +
\Big({1 \over 2} - {1 \over 2}\Big)   -   {1\over 3}  \Big]
\nonumber
\\[0.3cm]
+ g_{\alpha \beta} (\nabla^{\rho}\nabla_{\rho})  (\nabla^{\gamma}\Lambda_{\gamma})
\left[  \left( - { 1 \over 4 } + {1 \over 6 } + {1 \over 6 } - {1 \over 12 } \right)  +
\left(- {1 \over 4 } + {1 \over 6 } + {1 \over 6} - {1 \over 12 } \right)  \right]
\nonumber
\\[0.3cm]
+ \nabla_{\alpha}\nabla_{\beta} \; (\nabla^{\rho} \Lambda_{\rho})
\Big[\Big(-{1 \over 2} + {1\over 4} -
{1 \over 6} + {1 \over 12})  + \Big(-{1 \over 2} + {1\over 4} -
{1 \over 6} + {1 \over 12}\Big)
\nonumber \\
+ \Big({1 \over 2} -{1 \over 3} + {1 \over 6}\Big)  +
\Big({1 \over 2} -{1 \over 3} + {1 \over 6}\Big) \Big]
\nonumber
\\
+  \Big\{   -
{1 \over 2}  \nabla^{\rho}   [\nabla_{\alpha} , \nabla_{\rho }]_{-}  \Lambda_{\beta} -
{1 \over 2}  [\nabla^{\rho}, \nabla_{\alpha}  \nabla_{\beta}]_{-}  \Lambda_{\rho}
\nonumber
\end{eqnarray}
\begin{eqnarray}
 + {1 \over 6}  g_{\alpha \beta}  \nabla^{\rho}
 [\nabla^{\sigma}, \nabla_{\rho} ]_{-}  \Lambda_{\sigma}  +
{1 \over 6} g_{\alpha \beta}  [\nabla^{\rho}, \nabla^{\sigma}\nabla_{\sigma} ]_{-}
\Lambda_{\rho}  -
\nonumber
\\[0.3cm]
- {1 \over 6} \nabla_{\beta} [\nabla^{\sigma} , \nabla_{\alpha}]_{-} \Lambda_{\sigma} -
{1 \over 6} [\nabla_{\beta} , \nabla^{\sigma}\nabla_{\sigma} ]_{-} \Lambda_{\alpha} \Big\}
\nonumber \\
 +  \Big\{  -
{1 \over 2}  \nabla^{\rho} [\nabla_{\beta} , \nabla_{\rho }]_{-} \Lambda_{\alpha}  -
{1 \over 2} [\nabla^{\rho}, \nabla_{\beta}  \nabla_{\alpha}]_{-} \Lambda_{\rho}
\nonumber
\\[0.3cm]
 +   {1 \over 6}  g_{\beta \alpha}  \nabla^{\rho}
  [\nabla^{\sigma}, \nabla_{\rho} ]_{-}  \Lambda_{\sigma}  +
{1 \over 6} g_{\beta \alpha } [\nabla^{\rho}, \nabla^{\sigma}\nabla_{\sigma} ]_{-}
\Lambda_{\rho} \nonumber \\
-{1 \over 6} \nabla_{\alpha} [\nabla^{\sigma} , \nabla_{\beta}]_{-} \Lambda_{\sigma} -
{1 \over 6} [\nabla_{\alpha} , \nabla^{\sigma}\nabla_{\sigma} ]_{-} \Lambda_{\beta}  \Big\} +
\nonumber
\\[0.3cm]
 +  \Big\{  -  {1 \over 3} \nabla_{\alpha}  [\nabla^{\rho} ,  \nabla_{\beta} ]_{-}
\Lambda_{\rho}  -
  {1 \over 3}  [\nabla_{\alpha} ,  \nabla^{\rho} \nabla_{\rho}]_{-} \Lambda_{\beta} \Big\}
\nonumber \\
+ \Big\{ -  {1 \over 3} \nabla_{\beta} [\nabla^{\rho} ,  \nabla_{\alpha} ]_{-}
 \Lambda_{\rho} -
{1 \over 3}  [\nabla_{\beta} ,  \nabla^{\rho} \nabla_{\rho}]_{-}
\Lambda_{\alpha} \Big\}
\nonumber
\\[0.3cm]
 + {1 \over 6} g_{\alpha \beta} ( \nabla^{\rho} [\nabla^{\sigma} ,\nabla_{\rho}]_{-}
\Lambda_{\sigma}  +
 [\nabla^{\rho}, \nabla^{\sigma} \nabla_{\sigma}]_{-}
\Lambda_{\rho}  ) \}  .
\nonumber
\end{eqnarray}

\noindent
Calculating in series all commutators,
after simple calculation we will produce
\begin{eqnarray}
0 =
g_{\alpha \beta} \; \nabla_{\rho} \;( R^{\rho \sigma} \;  \Lambda_{\sigma})
+ \; \Lambda^{\sigma} \;
\Big[\; \nabla_{\rho} R^{\rho}_{\;\;\alpha\beta\sigma} \;  +\;
\nabla_{\rho} R^{\rho}_{\;\;\beta \alpha \sigma}\; \Big]
\nonumber\\
+ \;
(\nabla_{\rho} \Lambda_{\sigma})\;
\Big[\;
 R^{\rho\;\;\;\;\;\sigma}_{\;\;\alpha \beta}  \; + \;
 R^{\rho\;\;\;\;\;\sigma}_{\;\;\beta \alpha} \; \Big] \;
\nonumber
\\
- \; \Lambda^{\rho} \; \Big[ \;  \nabla_{\alpha} R_{\beta \rho} \; +
\nabla_{\beta} R_{\alpha  \rho} \;\Big]  \;
-{3\over 2} \;
\Big[\; R_{\beta}^{\;\;\rho} \; (\nabla_{\alpha}  \Lambda_{\rho})\; +\;
R_{\alpha}^{\;\;\rho} \; ( \nabla_{\beta}  \Lambda_{\rho})\;\Big]\;
\nonumber
\\
+\; {1 \over 2} \;  \Big[  R_{\beta}^{ \rho}  (\nabla_{\rho} \Lambda_{\alpha} ) +
     R_{\alpha}^{\;\;\rho} \; (\nabla_{\rho}\Lambda_{\beta})  \Big]
\; .
\label{2.15}
\end{eqnarray}

It must be noticed that contrary to the expectations the equation obtained
contains explicitly the curvature Riemann tensor.
It enters into Eq. (\ref{2.15}) in two combinations:
\begin{eqnarray}
\Lambda^{\sigma} \; (\nabla_{\rho} R^{\rho}_{\;\;\alpha\beta \sigma} \;  +\;
\nabla_{\rho} R^{\rho}_{\;\;\beta \alpha \sigma} )  ,\qquad
\label{2.16a}
(\nabla_{\rho} \Lambda_{\sigma}) \; [ R^{\rho\;\;\;\;\;\sigma}_{\;\;\alpha \beta}  \; + \;
  R^{\rho\;\;\;\;\;\sigma}_{\;\;\beta \alpha} \;]\;\; .
\end{eqnarray}

The curvature tensor in combination  (\ref{2.16a}) can be  readily escaped. To this end,
it suffices for  the Bianchi identity
\begin{eqnarray}
\nabla_{\gamma} R^{\rho}_{\;\;\alpha\; \beta \sigma} +
\nabla_{\sigma} R^{\rho}_{\;\;\alpha \;\gamma \beta} +
\nabla_{\beta} R^{\rho}_{\;\;\alpha \; \sigma \gamma }   =0,
\qquad
\nabla_{\rho} R^{\rho}_{\;\;\alpha\; \beta \sigma}  +
\nabla_{\sigma} R_{ \beta \alpha } - \nabla_{\beta} R_{\alpha  \sigma}   = 0 \;  .
\nonumber
\end{eqnarray}

\noindent
Thus,
\begin{eqnarray}
\nabla_{\rho} R^{\rho}_{\;\;\alpha\; \beta \sigma}
+ \nabla_{\rho} R^{\rho}_{\;\;\beta  \alpha \sigma}
= (\nabla_{\alpha}  R_{\beta  \sigma}  \; + \;  \nabla_{\beta} R_{\alpha  \sigma}  ) \; - \;
2 \nabla_{\sigma} R_{ \beta \alpha } \; .
\label{2.17c}
\end{eqnarray}

\noindent
With Eq. (\ref{2.17c}), Eq.   (\ref{2.15}) takes the form
\begin{eqnarray}
0 =
g_{\alpha \beta} \; \nabla_{\rho} \;( R^{\rho \sigma} \;  \Lambda_{\sigma}) \;
-
 2 \Lambda^{\sigma} \nabla_{\sigma} R_{\alpha \beta}
+ \;
(\nabla_{\rho} \Lambda_{\sigma})\;
[\;
 R^{\rho\;\;\;\;\;\sigma}_{\;\;\alpha \beta}   +
 R^{\rho\;\;\;\;\;\sigma}_{\;\;\beta \alpha}  ]
\nonumber
\\
-{3\over 2}
[  R_{\beta}^{\;\;\rho} \; (\nabla_{\alpha}  \Lambda_{\rho})  +
R_{\alpha}^{\;\;\rho}  ( \nabla_{\beta}  \Lambda_{\rho}) ]
+\; {1 \over 2} \;  [\; R_{\beta}^{\;\;\rho}\; (\nabla_{\rho} \Lambda_{\alpha} )\;+\;
     R_{\alpha}^{\;\;\rho} \; (\nabla_{\rho}\Lambda_{\beta}) \; ]\;
\; .
\label{2.18}
\end{eqnarray}

However, the curvature tensor still remains to  enter Eq. (\ref{2.18}).
And this means that in regions involving  curvature
the above massless spin-2 equation does not allow any gauge  principle.

Now we will show that in order to overcome such a difficulty the above starting equations
should be slightly altered. To this end, let us add special term (a not minimal gravitational
interaction term) into Eq. (\ref{2.1c}):
\begin{eqnarray}
{1 \over 2} \Big( \nabla^{\rho }  \Phi_{\rho  \alpha \beta }  +
\nabla^{\rho }  \Phi_{\rho  \beta  \alpha }  -   {1 \over 2}
g_{\alpha  \beta}(x)   \nabla^{\rho}  \Phi_{\rho
\sigma}^{\;\;\;\;\sigma }\; \Big)
\nonumber
\\[0.2cm]
+    \Big(  \nabla_{\alpha}  \Phi_{\beta}  +   \nabla_{\beta}  \Phi_{\alpha} -
{1 \over 2}  g_{\alpha \beta }(x)   \nabla^{\rho }  \Phi_{\rho} \Big)
= A    [  R^{\rho\;\;\;\;\;\sigma}_{\;\;\alpha \beta}  +
R^{\rho\;\;\;\;\;\sigma}_{\;\;\beta \alpha } ]  \Phi_{\rho \sigma}  .
\label{2.19}
\end{eqnarray}

\noindent Let us  show that at special parameter $A$  the theory of massless spin-2 particle
can be done satisfactory in the sense of the above gauge  principle. Indeed,
\begin{eqnarray}
A  R^{\rho\;\;\;\;\;\sigma}_{\;\;\alpha \beta}
\Phi^{(0)}_{\rho \sigma}
=
A   R^{\rho\;\;\;\;\;\sigma}_{\;\;\alpha \beta}
\Big( \nabla_{\rho} \Lambda_{\sigma}  + \nabla_{\sigma} \Lambda_{\rho}  -
{1 \over 2} g_{\rho \sigma}
\nabla^{\gamma} \Lambda_{\gamma} \Big)
\nonumber
\\
= A  \Big[   R^{\rho\;\;\;\;\;\sigma}_{\;\;\alpha \beta}  \nabla_{\rho} \Lambda_{\sigma} +
R^{\rho\;\;\;\;\;\sigma}_{\;\;\beta \alpha } \nabla_{\rho} \Lambda_{\sigma}  +
{1 \over  2}
R_{\alpha \beta} \nabla^{\gamma} \Lambda _{\gamma}   \Big]
\nonumber
\end{eqnarray}

\noindent and  therefore a contribution of that additional term into
(\ref{2.18}) is equal to
\begin{eqnarray}
A    [  R^{\rho\;\;\;\;\;\sigma}_{\;\;\alpha \beta}  +
R^{\rho\;\;\;\;\;\sigma}_{\;\;\beta \alpha } ]  \Phi^{(0)}_{\rho \sigma}
= 2A
[  R^{\rho\;\;\;\;\;\sigma}_{\;\;\alpha \beta}   +
R^{\rho\;\;\;\;\;\sigma}_{\;\;\beta \alpha }  ]
(\nabla_{\rho} \Lambda_{\sigma})  +
A  R_{\alpha \beta} \; (\nabla^{\gamma} \Lambda _{\gamma})  ]  .
\label{2.20b}
\end{eqnarray}

\noindent
So, instead of  Eq. (\ref{2.18})  we have
\begin{eqnarray}
2A  ( \nabla_{\rho} \Lambda_{\sigma})
[  R^{\rho\;\;\;\;\;\sigma}_{\;\;\alpha \beta}   +
R^{\rho\;\;\;\;\;\sigma}_{\;\;\beta \alpha }  ]   +
 A R_{\alpha \beta}  (\nabla^{\gamma} \Lambda _{\gamma}  )
\nonumber
\\[0.2cm]
=
g_{\alpha \beta}  \nabla_{\rho} ( R^{\rho \sigma}   \Lambda_{\sigma})
- 2 \Lambda^{\sigma} \nabla_{\sigma} R_{\alpha \beta}
+ (\nabla_{\rho} \Lambda_{\sigma})
[
R^{\rho\;\;\;\;\;\sigma}_{\;\;\alpha \beta}   +
 R^{\rho\;\;\;\;\;\sigma}_{\;\;\beta \alpha}  ]
\nonumber
\\[0.2cm]
-{3\over 2}
[ R_{\beta}^{\;\;\rho}  (\nabla_{\alpha}  \Lambda_{\rho})  +
R_{\alpha}^{\;\;\rho}  ( \nabla_{\beta}  \Lambda_{\rho}) ]
+  {1 \over 2}   [ R_{\beta}^{\;\;\rho}  (\nabla_{\rho} \Lambda_{\alpha} ) +
     R_{\alpha}^{\;\;\rho}  (\nabla_{\rho}\Lambda_{\beta})  ] .
\label{2.21a}
\end{eqnarray}

\noindent
Setting  $A = {1 \over 2}$, both terms with curvature tensor will be cancelled by each other:
\begin{eqnarray}
{1 \over 2}  R_{\alpha \beta}  (\nabla^{\gamma} \Lambda _{\gamma}  )
= g_{\alpha \beta}  \nabla_{\rho} ( R^{\rho \sigma}   \Lambda_{\sigma})
- 2 \Lambda^{\sigma} \nabla_{\sigma} R_{\alpha \beta}
\nonumber
\\[0.2cm]
-{3\over 2}
[ R_{\beta}^{\;\;\rho}  (\nabla_{\alpha}  \Lambda_{\rho}) +
R_{\alpha}^{\;\;\rho}  ( \nabla_{\beta}  \Lambda_{\rho})]
+
{1 \over 2}   [ R_{\beta}^{\;\;\rho} (\nabla_{\rho} \Lambda_{\alpha} ) +
     R_{\alpha}^{\;\;\rho}  (\nabla_{\rho}\Lambda_{\beta})  ] . \;\;\;
\label{2.21b}
\end{eqnarray}

Finally the obtained  relationship does not contain the curvature tensor and will turn into identity
at $R_{\alpha \beta}(x)=0$ which  was required.
So, the required system  is  which one changes Eq.~(\ref{2.1c})  by
\begin{eqnarray}
{1 \over 2}  \Big( \nabla^{\rho }  \Phi_{\rho  \alpha \beta }  +
\nabla^{\rho }  \Phi_{\rho  \beta  \alpha }  -   {1 \over 2}
g_{\alpha  \beta}(x)   \nabla^{\rho}  \Phi_{\rho
\sigma}^{\;\;\;\;\sigma }\; \Big)
\nonumber
\\[0.2cm]
+    \Big(  \nabla_{\alpha}  \Phi_{\beta}  +   \nabla_{\beta}  \Phi_{\alpha} -
{1 \over 2} g_{\alpha \beta }(x)   \nabla^{\rho }  \Phi_{\rho} \Big)
= {1 \over 2}   (  R^{\rho\;\;\;\;\;\sigma}_{\;\;\alpha \beta}  +
R^{\rho\;\;\;\;\;\sigma}_{\;\;\beta \alpha } )  \Phi_{\rho \sigma}  .
\label{2.12}
\end{eqnarray}


\end{document}